\documentclass[preprint,12pt]{elsarticle}
\usepackage[cp1251]{inputenc}
\usepackage{natbib}
\usepackage{amsmath}
\usepackage{amssymb}
\usepackage{textcomp}
\usepackage{graphicx}

\journal{Materials Today Communications}

\begin{document}

\begin{frontmatter}



\title{Metal--insulator transition in type II heterostructures\\
based on transition metal dichalcogenides}


\author{Pavel V. Ratnikov\footnote{Email address: ratnikov@lpi.ru}}

\address{A.M. Prokhorov General Physics Institute, Russian Academy of Sciences\\ Vavilova Street 38, 119991 Moscow, Russia}

\begin{abstract}
The problem of screening the Coulomb interaction between charge carriers in type II heterostructures based on transition metal dichalcogenides is Analytically solved. At a sufficiently high density of charge carriers, the density dependence of the interlayer exciton energy is obtained. The energy of the interlayer exciton tends to zero in the metal--insulator transition point. The presented scheme of calculations makes it possible to find the temperature dependence of the density of this transition.
\end{abstract}


\end{frontmatter}


\section{Introduction}
Recently, interest has increased in the metal--insulator transition in van der Waals (vdW) heterostructures composed of various layers of transition metal dichalcogenides (TMDs). Both theoretically and experimentally, bilayers formed by superimposing one layer of monomolecular thickness (monolayer) of TMD on another are intensively studied. Such bilayers have a finite (though small) twist angle of the crystal lattice of one monolayer with respect to the crystal lattice of another monolayer. As a result, the moir\'{e}-type structure appears, which creates an additional periodic potential for charge carriers. In fact, it manifests itself as a two-dimensional (2D) superlattice. A feature of such twisted bilayers is the appearance of flat bands in the energy spectrum of charge carriers \cite{Zhang2020}.

Observation of the influence of flat bands on the transport characteristics of bilayers requires very low temperatures (below helium). For example, experiments with tungsten diselenide were carried out at 1.8 K \cite{Wang2020}. This is due to the small energy gaps that separate flat bands from ordinary parabolic bands at the $K$ points of the Brillouin zone (apparently, these gaps are tenths of meV).

Heterobilayers (bilayers of monolayers of different materials) have half-filled narrow moir\'{e} bands. The parameters of the moir\'{e} superlattice are determined by the materials that make up the heterobilayer and the twist angle. The charge gap in the moir\'{e} superlattice can vanish depending on the ratio of the interaction energy to the kinetic energy. This indicates a metal--insulator transition \cite{Morales-Duran2021}.

The results of an experimental study of the metal--insulator transition in the heterobilayer WSe$_2$/MoSe$_2$ are presented in the work \cite{Wang2019}. Having the type II contact, this system makes it possible to observe with increasing in the density of electron-hole ($e$-$h$) pairs the transition from interlayer excitons to a charge-separated $e$-$h$ plasma (holes are in WSe$_2$, and electrons are in MoSe$_2$). The density of this transition $3\times10^{12}$ cm$^{-2}$ was obtained from the quenching of the exciton line in the photoluminescence spectrum.

The discovery of a high-temperature electron-hole liquid (EHL) in TMD monolayers \cite{Yu2019, Arp2019} convincingly confirmed that they are ideal systems for studying electronic, optoelectronic, and quantum phenomena. Quite recently, we have predicted the formation of the charge-separated EHL in the type II TMD heterostructures \cite{Ratnikov2022}. In our opinion, the appearance of the broad EHL line in the photoluminescence spectrum of such heterostructures promotes greater absorption of light and an increase in the efficiency of solar cells.

In the present work, we theoretically study the metal--insulator transition in the type II TMD heterostructures in the region of a sufficiently high $e$-$h$ pair density at an arbitrary temperature (including close to room temperature). The appearance of moir\'{e} gaps in the energy spectrum of charge carriers becomes insignificant for the consideration of this transition. First, we find the screened potential within the framework of the electrodynamics of continuous media and then we calculate variationally the interlayer exciton energy. The condition of vanishing of the latter is taken by us as the transition criterion.

\section{Preliminary remarks}
We consider the type II TMD heterostructures from the point of view of continuum electrodynamics. They are thin films of thickness $d$, consisting of two layers, the contact between which is assumed to be continuous. Let there be holes in the first layer and electrons in the second layer. Physically, the region occupied by holes (hole layer) is separated by the vdW gap from the region where electrons are located (electron layer). Since we adhere to the framework of the electrodynamics of continuous media, we first assume that holes and electrons are located in an arbitrary place, respectively, of the first and second layers, and then we take into account their separation by the vdW gap, when averaging over their position across the heterostructure plane is carried out.

As a criterion for the metal--insulator transition, we take the vanishing of the interlayer exciton energy:
\begin{equation}\label{CriterionMIT}
E_\text{ex}(n_{dm})=0.
\end{equation}
Here, $n_{dm}$ is the density of this transition.

We note that to determine the $E_\text{ex}(n)$ dependence, it is necessary to take into account the dynamic screening of the Coulomb interaction between an electron and a hole that make up the interlayer exciton. Its characteristic frequency is determined by the interlayer exciton binding energy. Since we are interested in the region of densities $n\simeq n_{dm}$ and the characteristic frequencies tend to zero according to criterion \eqref{CriterionMIT}, we can restrict ourselves to static screening and solve the problem within the framework of electrostatics.

We also believe that the $e$-$h$ pair density $n$ is large enough for the average distance between an electron and a hole in an interlayer exciton (the Bohr radius $a_B$) to be much larger than the average distance between particles in the system $\bar{r}$, but at the same time to remain large compared to the film thickness:
\begin{equation}\label{Conditions}
a_B\gg\bar{r}\gg d.
\end{equation}
The first condition makes it possible to consider the potential at large distances when solving the problem for the interlayer exciton without going into the details of its behavior at small distances. The second condition determines the 2D nature of the problem.

\section{Screened electrostatic potential}
\begin{figure}[b!]
\begin{center}
\includegraphics[width=0.8\textwidth]{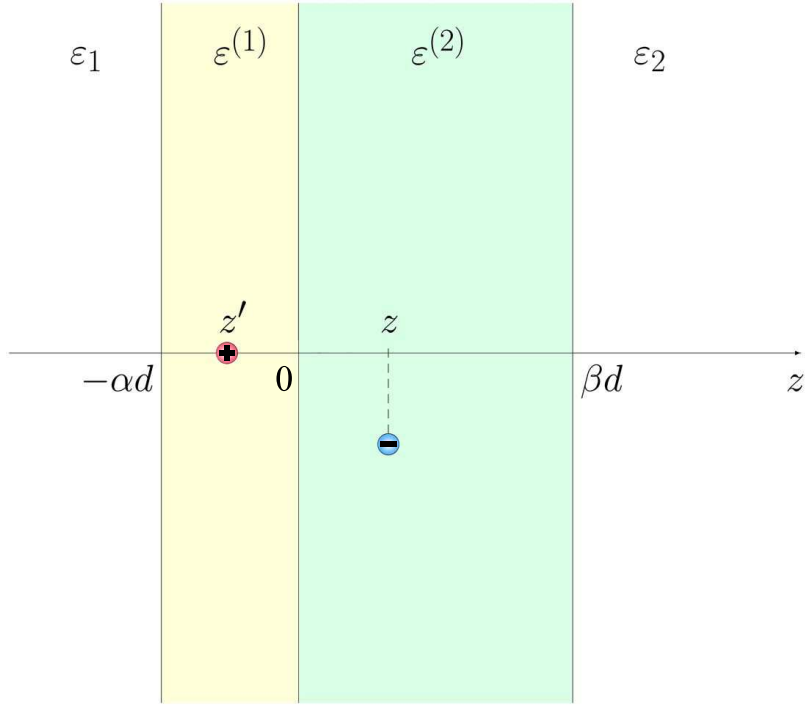}
\end{center}
\caption{Schematic representation of the system under consideration.}
\label{f1}
\end{figure}

Let the $z$ axis be directed along the normal to the heterostructure plane. The plane $z=0$ is the boundary between the two semiconductor layers that make up the heterostructure (see Fig. \ref{f1}). The first layer has a thickness $\alpha d$ and the second layer has a thickness $\beta d$, $\alpha+\beta=1$. The dielectric environment occupies the regions $z<-\alpha d$ (the medium with permittivity $\varepsilon_1$) and $z>\beta d$ (the medium with permittivity $\varepsilon_2$). The permittivities of the semiconductor layer materials are $\varepsilon^{(1)}$ (the first layer) and $\varepsilon^{(2)}$ (the second layer).

We assume as a trial charge a hole with coordinate $z^\prime$ on the $z$ axis. We are interested in the value of the electrostatic potential at the point $(\boldsymbol{\rho},\,z)$ inside the second layer, created by the trial charge, taking into account its screening from other holes in the first layer and electrons in the second layer. The system as a whole is electrically neutral one.

The screened electrostatic potential satisfies the equations in four areas:
\begin{equation}\label{SystemEquations1}
\begin{split}
\triangle_\mathbf{r}\varphi_1(\mathbf{r},\,\mathbf{r}^\prime)&=0,~~~z<-\alpha d,\\
\triangle_\mathbf{r}\varphi^{(1)}(\mathbf{r},\,\mathbf{r}^\prime)&=-\frac{4\pi e}{\varepsilon^{(1)}}\left[\delta(\mathbf{r}-\mathbf{r}^\prime)-\Delta n^{(1)}_\text{3D}(\mathbf{r},\,\mathbf{r}^\prime)\right],~~~-\alpha d<z<0,\\
\triangle_\mathbf{r}\varphi^{(2)}(\mathbf{r},\,\mathbf{r}^\prime)&=\frac{4\pi e}{\varepsilon^{(2)}}\Delta n^{(2)}_\text{3D}(\mathbf{r},\,\mathbf{r}^\prime),~~~0<z<\beta d,\\
\triangle_\mathbf{r}\varphi_2(\mathbf{r},\,\mathbf{r}^\prime)&=0,~~~z>\beta d.
\end{split}
\end{equation}

Under conditions of thermal equilibrium, the density of induced charges, $\Delta n^{(1)}_\text{3D}(\mathbf{r},\,\mathbf{r}^\prime)$ for holes in the first layer and $\Delta n^{(2)}_\text{3D}(\mathbf{r},\,\mathbf{r}^\prime)$ for electrons in the second layer, is a function of the electrochemical potential
\begin{equation}
\mu=\mu_e+\mu_h+e\varphi,
\end{equation}
where $\mu_{e(h)}$ is the chemical potential of electrons (holes) in the absence of the trial charge,
\begin{equation}\label{ChemPotential}
\mu_{e(h)}=T\ln\left(e^{E^{e(h)}_F/T}-1\right),
\end{equation}
$E^{e(h)}_F$ is the Fermi energy of electrons (holes), $E^e_F=\frac{2\pi n}{(1+\sigma)\nu_e}$ and $E^h_F=\frac{2\pi\sigma n}{(1+\sigma)\nu_h}$, $\sigma=m_{e2}/m_{h1}$ is the ratio of the mass of an electron in the second layer to the mass of a hole in the first layer, $\nu_e$ and $\nu_h$ are the numbers of electron and hole valleys in the second and first layers, respectively, $n$ is the surface density of $e$-$h$ pairs. The units of measurement for energy (temperature) and distance are the binding energy and the Bohr radius of a 2D exciton in the zero density limit.

Also, due to conditions \eqref{Conditions}, the density of induced charges $\Delta n^{(1)}_\text{3D}$ and $\Delta n^{(2)}_\text{3D}$ as functions of coordinates are close, differing from each other by a factor depending on the $e$-$h$ pair density $n$ and temperature $T$. We are interested in the behavior of the screened electrostatic potential at distances $\rho\sim a_B$. The interlayer exciton binding energy in a dense system of $e$-$h$ pairs is small compared to the binding energy in the zero density limit $|E_\text{ex}(0)|$. This means that the energy of their interaction averaged over the positions of an electron and a hole in the $xy$ plane (in modulus) is $\langle e\varphi\rangle_{xy}\simeq|E_\text{ex}(n)|\ll|E_\text{ex}(0)|$. On the other hand, the Fermi energy is $E_F=E^e_F+E^h_F\simeq|E_\text{ex}(0)|$ and, consequently, $\langle e\varphi\rangle_{xy}\ll E_F$. For the characteristic values of $n$ and $T$ for the coexistence of EHL and exciton gas ($e$-$h$ plasma) $T\lesssim\frac{1}{10}E_F$ and we are in the quantum region. The density of induced charges can be expanded into a series in powers of $\varphi$. We restrict ourselves, as in the work \cite{Rytova1967}, to the first linear term, taking into account expression \eqref{ChemPotential}
\begin{equation}\label{IndChargeDensity}
\Delta n^{(1,2)}_\text{3D}=\frac{1}{\delta d}\frac{\partial n}{\partial\mu_{h,e}}e\varphi,
\end{equation}
where
\begin{equation}\label{dn/dmu}
\begin{split}
\frac{\partial n}{\partial\mu_e}&=\frac{(1+\sigma)\nu_e}{2\pi}\left(1-e^{-2\pi n/(1+\sigma)\nu_eT}\right),\\
\frac{\partial n}{\partial\mu_h}&=\frac{(1+\sigma)\nu_h}{2\pi\sigma}\left(1-e^{-2\pi\sigma n/(1+\sigma)\nu_hT}\right).
\end{split}
\end{equation}
Here we took into account that $n_\text{3D}=n/\delta d$, $\delta d$ is the thickness of the electron (hole) layer.

Let's make a 2D Fourier transform for the system of equations \eqref{SystemEquations1}
\begin{equation}\label{SystemEquationsFourier}
\begin{split}
&\frac{\partial^2\varphi_1(\mathbf{k};\,z,\,z^\prime)}{\partial z^2}-k^2\varphi_1(\mathbf{k};\,z,\,z^\prime)=0,\\
&\frac{\partial^2\varphi^{(1)}(\mathbf{k};\,z,\,z^\prime)}{\partial z^2}-\widetilde{k}^2\varphi^{(1)}(\mathbf{k};\,z,\,z^\prime)=-\frac{4\pi e}{\varepsilon^{(1)}}\delta(z-z^\prime),\\
&\frac{\partial^2\varphi^{(2)}(\mathbf{k};\,z,\,z^\prime)}{\partial z^2}-\overline{k}^2\varphi^{(2)}(\mathbf{k};\,z,\,z^\prime)=0,\\
&\frac{\partial^2\varphi_2(\mathbf{k};\,z,\,z^\prime)}{\partial z^2}-k^2\varphi_2(\mathbf{k};\,z,\,z^\prime)=0,
\end{split}
\end{equation}
where the following notations are introduced
\begin{equation*}
\widetilde{k}^2=k^2+\frac{2e^2}{\varepsilon^{(1)}\delta d}(1+\sigma^{-1})\nu_h\left(1-e^{-2\pi\sigma n/(1+\sigma)\nu_hT}\right),~\overline{k}^2=k^2+\frac{2e^2}{\varepsilon^{(2)}\delta d}(1+\sigma)\nu_e\left(1-e^{-2\pi n/(1+\sigma)\nu_eT}\right).
\end{equation*}
General solutions of equations \eqref{SystemEquationsFourier} are
\begin{equation}\label{GeneralSolutions}
\begin{split}
\varphi_1(\mathbf{k};\,z,\,z^\prime)&=A_1e^{kz},\\
\varphi^{(1)}(\mathbf{k};\,z,\,z^\prime)&=A^{(1)}e^{\widetilde{k}z}+B^{(1)}e^{-\widetilde{k}z}+\frac{2\pi e}{\varepsilon^{(1)}\widetilde{k}}e^{-\widetilde{k}|z-z^\prime|},\\
\varphi^{(2)}(\mathbf{k};\,z,\,z^\prime)&=A^{(2)}e^{\overline{k}z}+B^{(2)}e^{-\overline{k}z},\\
\varphi_2(\mathbf{k};\,z,\,z^\prime)&=B_2e^{-kz}.
\end{split}
\end{equation}
We apply boundary conditions at $z=-\alpha d$, $z=0$, and $z=\beta d$ to the solutions \eqref{GeneralSolutions}
\begin{equation}\label{SystemEquations2}
\begin{split}
&A_1e^{-\alpha kd}=A^{(1)}e^{-\alpha\widetilde{k}d}+B^{(1)}e^{\alpha\widetilde{k}d}+\frac{2\pi e}{\varepsilon^{(1)}\widetilde{k}}e^{-\widetilde{k}(\alpha d+z^\prime)},\\
&A_1\varepsilon_1ke^{-\alpha kd}=A^{(1)}\varepsilon^{(1)}\widetilde{k}e^{-\alpha\widetilde{k}d}-B^{(1)}\varepsilon^{(1)}\widetilde{k}e^{\alpha\widetilde{k}d}+2\pi e\cdot e^{-\widetilde{k}(\alpha d+z^\prime)},\\
&A^{(1)}+B^{(1)}+\frac{2\pi e}{\varepsilon^{(1)}\widetilde{k}}e^{\widetilde{k}z^\prime}=A^{(2)}+B^{(2)},\\
&A^{(1)}\varepsilon^{(1)}\widetilde{k}-B^{(1)}\varepsilon^{(1)}\widetilde{k}-2\pi e\cdot e^{\widetilde{k}z^\prime}=A^{(2)}\varepsilon^{(2)}\overline{k}-B^{(2)}\varepsilon^{(2)}\overline{k},\\
&A^{(2)}e^{\beta\overline{k}d}+B^{(2)}e^{-\beta\overline{k}d}=B_2e^{-\beta kd},\\
&A^{(2)}\varepsilon^{(2)}\overline{k}e^{\beta\overline{k}d}-B^{(2)}\varepsilon^{(2)}\overline{k}e^{-\beta\overline{k}d}=-B_2\varepsilon_2ke^{-\beta kd}.
\end{split}
\end{equation}
At extracting the modulus, we take into account that $-\alpha d<z^\prime<0$.

The value of the Fourier component of the screened electrostatic potential in the second layer is
\begin{equation}\label{FourierComponentScreenedPotential}
\varphi^{(2)}(\mathbf{k};\,z,\,z^\prime)=\frac{4\pi e\cdot\cosh\left(\widetilde{k}(\alpha d+z^\prime)+\eta_1\right)\cosh\left(\overline{k}(\beta d-z)+\eta_2\right)}
{\varepsilon^{(1)}\widetilde{k}\sinh\left(\alpha\widetilde{k}d+\eta_1\right)\cosh\left(\beta\overline{k}d+\eta_2\right)+\varepsilon^{(2)}\overline{k}\cosh\left(\alpha\widetilde{k}d+\eta_1\right)\sinh\left(\beta\overline{k}d+\eta_2\right)},
\end{equation}
where
\begin{equation*}
\eta_1=\frac{1}{2}\ln\frac{\varepsilon^{(1)}\widetilde{k}+\varepsilon_1k}{\varepsilon^{(1)}\widetilde{k}-\varepsilon_1k},~\eta_2=\frac{1}{2}\ln\frac{\varepsilon^{(2)}\overline{k}+\varepsilon_2k}{\varepsilon^{(2)}\overline{k}-\varepsilon_2k}.
\end{equation*}

The trial charge (hole) that creates this potential is in the first layer, $-\alpha d<z^\prime<-\alpha d+\delta d$; the electron on which it acts is in the second layer, $\alpha d-\delta d<z<\alpha d$. Between them there is the vdW gap $\Delta d=2(\alpha d-\delta d)$. In the case of a monolayer/monolayer heterostructure, there is only one vdW gap, and in the case of a monolayer/bilayer heterostructure, there are two vdW gap, but the electrons in the bilayer, being attracted to the holes in the monolayer, flock to the lower layer in the bilayer, and the vdW gap remains the same.

Now we average $\varphi^{(2)}$ over the positions of the trial charge $z^\prime$ and the charge on which it acts $z$, assuming that their distribution across the hole and electron layers, respectively, is given by the wave functions of ``transverse motion''
\begin{equation*}
\psi_1(z^\prime)=\sqrt{\frac{2}{\delta d}}\cos\left(\pi\frac{z^\prime+\alpha d-\delta d/2}{\delta d}\right),~\psi_2(z)=\sqrt{\frac{2}{\delta d}}\cos\left(\pi\frac{z-\alpha d+\delta d/2}{\delta d}\right);
\end{equation*}
\begin{equation}\label{AverageFourierComponentScreenedPotential}
\begin{split}
&\overline{\varphi}^{(2)}(\mathbf{k})\equiv\int\limits_{-\alpha d}^{-\alpha d+\delta d}dz^\prime\int\limits_{\alpha d-\delta d}^{\alpha d}dz\varphi^{(2)}(\mathbf{k};\,z,\,z^\prime)\left|\psi_1(z^\prime)\right|^2\left|\psi_2(z)\right|^2\\
=&\frac{2^8\pi^5e\cosh\left(\frac{1}{2}\widetilde{k}\delta d+\eta_1\right)\sinh\left(\frac{1}{2}\widetilde{k}\delta d\right)\cosh\left(\overline{k}\left((\beta-\alpha)d+\frac{1}{2}\delta d\right)+\eta_2\right)\sinh\left(\frac{1}{2}\overline{k}\delta d\right)}{\left[\varepsilon^{(1)}\widetilde{k}\sinh\left(\alpha\widetilde{k}d+\eta_1\right)\cosh\left(\beta\overline{k}d+\eta_2\right)
+\varepsilon^{(2)}\overline{k}\cosh\left(\alpha\widetilde{k}d+\eta_1\right)\sinh\left(\beta\overline{k}d+\eta_2\right)\right]\widetilde{k}\overline{k}\delta d^2}\\
&\times\frac{1}{\left(\widetilde{k}^2\delta d^2+4\pi^2\right)\left(\overline{k}^2\delta d^2+4\pi^2\right)}.
\end{split}
\end{equation}

Expression \eqref{AverageFourierComponentScreenedPotential} is the desired answer. It gives the interaction energy $-e\overline{\varphi}^{(2)}(\boldsymbol{\rho})$ (after the inverse Fourier transform) of an electron in the second layer and a hole in the first layer of the heterostructure under consideration. In view of the cumbersomeness of the expression $\overline{\varphi}^{(2)}(\boldsymbol{\rho})$, it is convenient to work with expression \eqref{AverageFourierComponentScreenedPotential} while remaining in $\mathbf{k}$ space.

\section{Interlayer exciton binding energy in the $e$-$h$ pair dense system}
In order to determine the density dependence of the interlayer exciton binding energy, we solve the Schr\"{o}dinger equation in $\mathbf{k}$ space by the variational method:
\begin{equation}\label{SchrodengerEquation}
\frac{\mathbf{q}^2}{2m}\psi(\mathbf{q})-e\int\frac{d^2k}{(2\pi)^2}\overline{\varphi}^{(2)}(\mathbf{k})\psi(\mathbf{k}-\mathbf{q})=E_\text{ex}\psi(\mathbf{q}),
\end{equation}
where $m=m_{e2}m_{h1}/(m_{e2}+m_{h1})$ is the reduced mass of an electron in the second layer and a hole in the first layer.

The trial wave function is chosen as the Fourier transform of the exponentially decreasing function ($a$ is the variational parameter)
\begin{equation}\label{TrialWaveFunction}
\widetilde{\psi}(\mathbf{q})=\frac{\sqrt{8\pi}a^2}{(a^2+q^2)^{3/2}}.
\end{equation}

We multiply \eqref{SchrodengerEquation} by $\widetilde{\psi}^*(\mathbf{q})$ and integrate over $\mathbf{q}$
\begin{equation}\label{EnergyExciton}
E_\text{ex}=\frac{a^2}{2m}-\frac{ea^4}{\pi^2}\int\limits_0^\infty kdk\int\limits_0^\infty qdq\int\limits_0^{2\pi}d\phi\frac{\overline{\varphi}^{(2)}(\mathbf{k})}{(a^2+q^2)^{3/2}(a^2+k^2+q^2-2kq\cos\phi)^{3/2}}.
\end{equation}
The interlayer exciton binding energy is obtained by minimizing the expression \eqref{EnergyExciton} with respect to the variational parameter.

Density and temperature go only to the arguments of the exponent~in~\eqref{dn/dmu}. For given values of the remaining parameters of the heterostructure, the ratio of $n$ to $T$ is a fixed value corresponding the criterion \eqref{CriterionMIT}. Whence it follows that
\begin{equation}\label{ndmT}
n_{dm}\propto T.
\end{equation}

However, it should be noted that the process of transition of the exciton gas into the $e$-$h$ plasma begins before approaching $n_{dm}$ due to the thermal ionization of excitons. There is a fairly wide range of densities adjacent to $n_{dm}$, where the $e$-$h$ system is a mixture of the exciton gas and $e$-$h$ plasma. Such a state is characterized by the degree of ionization $\xi=n_{pl}/n$ ($n_{pl}$ is the $e$-$h$ plasma density). At the point $n=n_{dm}$, $\xi=1$, and far from it, when $n\ll n_{dm}$, $\xi\ll1$. Since we are in the quantum region, the Saha equation turns out to be inapplicable for the calculation of $\xi$. Quantum-mechanical calculations should be carried out. The most convenient is the mathematical apparatus of spectral functions, implemented in the ionization equilibrium theory \cite{Steinhoff2017}. We do not carry out such calculations, since they are far beyond the scope of this work.

\begin{figure}[b!]
\begin{center}
\includegraphics[width=0.8\textwidth]{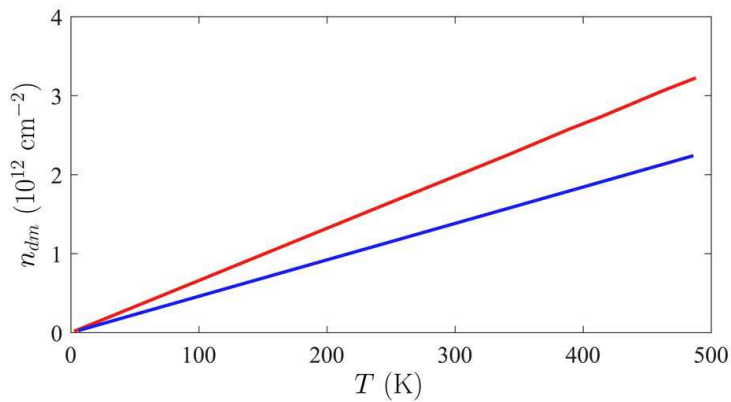}
\end{center}
\caption{Temperature dependence of the metal--insulator transition density for the heterostructure WSe$_2$/MoSe$_2$. The red curve is for the monolayer/monolayer, and the blue curve is for the monolayer/bilayer. The parameter values are taken from the works cited in \cite{Ratnikov2022}.}
\label{f2}
\end{figure}

Fig. \ref{f2} clearly demonstrates the linear dependence of the metal--insulator transition density on temperature (the heterostructure WSe$_2$/MoSe$_2$ on the SiO$_2$ substrate is taken as an example). As the number of monolayers increases, the slope of the straight line decreases. This is due to stronger screening of the Coulomb interaction in thicker heterostructures.

A decrease in temperature according to \eqref{ndmT} implies a decrease in density. In the limit $T\rightarrow0$ $n_{dm}\rightarrow0$, while our previous calculations for TMD heterostructures \cite{Pekh2021} showed that $n_{dm}\neq0$ in this case. This indicates the inapplicability of the approach presented here to the region of very low temperatures. The system is no longer dense enough (the density of induced charges $\Delta n^{(1,2)}_\text{3D}$ turns out to be comparable with the initial density of charge carriers $n$).

\section{Results and discussion}
In this work, we found the screened electrostatic potential of a charge in the thin film consisting of two films. This model problem made it possible to find the interaction energy of an electron and a hole in type II TMD heterostructures, taking into account the screening of the Coulomb interaction between them. Using the variational method, we calculated the interlayer exciton binding energy in such heterostructures at a finite charge carrier density. Its vanishing determines the metal--insulator transition density $n_{dm}$. We have obtained the linear temperature dependence $n_{dm}(T)$.

The density $n_{dm}$ in the region $T\simeq T_c$ is overestimated, $n_{dm}\simeq2n_c$ ($T_c$ and $n_c$ are the critical temperature and density of the gas--liquid transition). Strictly speaking, this means that the metal--insulator transition occurs in the liquid phase, while, as a rule, it occurs in the gas phase. But the accuracy of density calculations is such that an error of 2 times can be acceptable. On the other hand, such a discrepancy may indicate an insufficient consideration of the screening of the Coulomb interaction (lower densities are required for its screening). This can be corrected by adding to~\eqref{IndChargeDensity} the following terms of the expansion in powers of $e\varphi$. Then, the system of equations on $\varphi$ becomes a system of non-linear equations. It can only be solved numerically. For example, it can be solved by iterations, taking the solution obtained here as a zero approximation.

~

\textbf{Acknowledgments}

The work was supported by the Foundation for the Advancement of Theoretical Physics and Mathematics ``BASIS'' (the project no. 20-1-3-68-1).

\end{document}